\documentclass[12pt]{article}
\usepackage{amsmath}
\usepackage[pdftex]{graphicx}

\begin{document}

\begin{center}
\section*{Risk Model Based on General Compound Hawkes Process}
{\sc Anatoliy Swishchuk}\footnote{University of Calgary, Calgary, Canada} \footnote{The author wishes to thank NSERC for continuing support}

\end{center}

\hspace{1cm}

{\bf Abstract:} In this paper, we introduce a new model for the risk process based on general compound Hawkes process (GCHP) for the arrival of claims. We call it risk model based on general compound Hawkes process (RMGCHP). The Law of Large Numbers (LLN) and the Functional Central Limit Theorem (FCLT) are proved. We also study the main properties of this new risk model,  net profit condition, premium principle and ruin time (including ultimate ruin time) applying the LLN and FCLT for the RMGCHP. We show, as applications of our results,  similar results for risk model based on compound Hawkes process (RMCHP) and apply them to the classical risk model based on compound Poisson process (RMCPP). 

\hspace{1cm}

{\bf Keywords}: Hawkes process; general compound Hawkes process; risk model; net profit condition; premium principle; ruin time; ultimate ruin time; LLN; FCLT

\section{Introduction}

The Hawkes process (Hawkes (1971)) is a simple point process that has self-exciting property, clustering effect and long memory. 

It has been widely applied in seismology, neuroscience, DNA modelling and many other fields, including finance (Embrechts, Liniger and Lin (2011)) and insurance (Stabile et al. (2010)).

In this paper, we introduce a new model for the risk process, based on general compound Hawkes process (GCHP) for the arrival of claims. We call it {\it risk model based on general compound Hawkes process} (RMGCHP). To the best of the author's knowledge, this risk model is the most general relaying on the existing literature. Compound Hawkes process and risk model based on it was introduced in Stabile et al. (2010). 

In comparison to simple Poisson arrival of claims, GCHP model accounts for the risk of contagion and clustering of claims.

We note, that Stabile \& Torrisi (2010) were the first who replaced Poisson process by a simple Hawkes process in studying the classical problem of the probability of ruin. Dassios and Zhao (2011) considered the same ruin problem using marked mutually-exciting process (dynamic contagion process).

Jang \& Dassios (2012) implement Dassios \& Zhao (2011) to calculate insurance premiums and suggest higher premiums should be set up in general across different insurance product lines. 
\
Semi-Markov risk processes and their optimal control and stability were first introduced in Swishchuk \& Goncharova (1998) and studied and developed in Swishchuk (2000).

Compound Hawkes processes were applied to Limit Order Books in Swishchuk, Chavez-Casillas, Elliott and Remillard (2017). General compound Hawkes processes have also been applied to LOB in Swishchuk (2017). The general compound Hawkes process was first introduced in Swishchuk (2017) to model a risk process in insurance.

The paper is organized as follows. Section 2 is devoted to the description of Hawkes process. Section 3 contains Law of Large Numbers (LLN) and Functional Central Limit Theorem (FCLT) for RMGCHP. Section 4 contains applications of LLN and FCLT, including net profit condition, premium principle, ruin and ultimate ruin probabilities, and the probability density function of the time to ruin for RMGCHP. Section 5 describes applications of the results from Section 4 to the risk model based on compound Hawkes process (RMCHP). Section 5 contain the applications of the results from Section 5 to the classical risk model based on compound Poisson process (RMCPP), just for the completeness of the presentation. And Section 6 concludes the paper and highlights future work.

\section{Hawkes, General Compound Hawkes Process (GCHP) and Risk Model based on GCHP}

In this section we introduce Hawkes and general compound Hawkes processes and give some of their properties. We also introduce the risk model based on GCHP.

\subsection{Hawkes Process}

{\bf Definition 1 (Counting Process)}. A counting process is a stochastic process $N(t), t\geq 0,$ taking positive integer  values and satisfying: $N(0)=0.$ It is almost surely finite, and is a right-continuous step function with increments of size $+1.$ (See, e.g., Daley and Vere-Jones (1988)).

Denote by ${\cal F}^N(t), t\geq 0,$ the history of the arrivals up to time $t,$ that is, $\{{\cal F}^N(t), t\geq 0\},$ is a filtration, (an increasing sequence of $\sigma$-algebras).

A counting process $N(t)$ can be interpreted as a cumulative count of the number of arrivals into a system up to the current time $t.$ 

The counting process can also be characterized by the sequence of random arrival times $(T_1,T_2,...)$ at which the counting process $N(t)$ has jumped. The process defined by these arrival times is called a point process.

{\bf Definition 2 (Point Process).} If a sequence of random variables $(T_1,T_2,...),$ taking values in $[0,+\infty),$ has $P(0\leq T_1\leq T_2\leq...)=1,$ and the number of points in a bounded region is almost surely finite, then, $(T_1,T_2,...)$ is called a point process. (See, e.g., Daley, D.J. and Vere-Jones, D. (1988)).

{\bf Definition 3 (Conditional Intensity Function).} Consider a counting process $N(t)$ with associated histories ${\cal F}^N(t), t\geq 0.$ If a non-negative function $\lambda(t)$ exists such that 
$$
\lambda(t)=\lim_{h\to 0}\frac{E[N(t+h)-N(t)|{\cal F}^N(t)]}{h},
\eqno{(1)}
$$
then it is called the conditional intensity function of $N(t).$ We note, that sometimes this function is called the hazard function.

{\bf Definition 4 (One-dimensional Hawkes Process) (Hawkes (1971))}. The one-dimensional Hawkes process is a point point process $N(t)$ which is characterized by its intensity $\lambda(t)$ with respect to its natural filtration:
$$
\lambda (t)=\lambda+\int_{0}^t\mu(t-s)dN(s),
\eqno{(2)}
$$
where $\lambda>0,$ and the response function $\mu(t)$ is a positive function and satisfies $\int_0^{+\infty}\mu(s)ds<1.$

The constant $\lambda$ is called the background intensity and the function $\mu(t)$ is sometimes also called theexcitation function. We suppose that $\mu(t)\not= 0$ to avoid the trivial case, which is, a homogeneous Poisson process. Thus, the Hawkes process is a non-Markovian extension of the Poisson process.

The interpretation of equation (2) is that the events occur according to an intensity with a background intensity $\lambda$ which increases by $\mu(0)$ at each new event then decays back to the background intensity value according to the function $\mu(t).$ Choosing $\mu(0)>0$ leads to a jolt in the intensity at each new event, and this feature is often called a self-exciting feature, in other words, because an arrival causes the conditional intensity function $\lambda(t)$ in (1)-(2) to increase then the process is said to be self-exciting.

With respect to definitions of $\lambda(t)$ in (1) and $N(t)$ (2), it follows that
$$
P(N(t+h)-N(t)=m|{\cal F}^N(t))=\left\{
\begin{array}{rcl}
\lambda(t)h+o(h), &&m=1\\
o(h),&&m>1\\
1-\lambda(t)h+o(h),&&m=0.\\
\end{array}
\right.
$$
We should mention that the conditional intensity function $\lambda(t)$ in (1)-(2) can be associated with the compensator $\Lambda(t)$ of the counting process $N(t),$ that is:
$$
\Lambda(t)=\int_0^t\lambda(s)ds.
\eqno{(3)}
$$

Thus, $\Lambda(t)$ is the unique ${\cal F}^N(t), t\geq 0,$ predictable function, with $\Lambda(0)=0,$ and is non-decreasing, such that 
$$
N(t)=M(t)+\Lambda(t)\quad a.s.,
$$
where $M(t)$ is an ${\cal F}^N(t), t\geq 0,$ local martingale (This is the Doob-Meyer decomposition  of $N.$)

A common choice for the function $\mu(t)$ in (2) is one of exponential decay: 
$$
\mu(t)=\alpha e^{-\beta t},
\eqno{(4)}
$$ 
with parameters $\alpha,\beta>0.$ In this case the Hawkes process is called the Hawkes process with exponentially decaying intensity.

Thus, the equation (2) becomes
$$
\lambda (t)=\lambda+\int_{0}^t\alpha e^{-\beta (t-s)}dN(s),
\eqno{(5)}
$$
We note, that in the case of (4), the process $(N(t),\lambda(t))$ is a continuous-time Markov process, which is not the case for the choice (2). 

With some initial condition $\lambda(0)=\lambda_0,$ the conditional density $\lambda(t)$ in (5) with the exponential decay in (4) satisfies the SDE
$$
d\lambda(t)=\beta(\lambda-\lambda(t))dt+\alpha dN(t), \quad t\geq 0,
$$
which can be solved (using stochastic calculus) as
$$
\lambda (t)=e^{-\beta t}(\lambda_0-\lambda)+\lambda+\int_{0}^t\alpha e^{-\beta (t-s)}dN(s),
$$
which is an extension of (5).

 Another choice for $\mu(t)$ is a power law function:
$$
\lambda (t)=\lambda+\int_{0}^t\frac{k}{(c+(t-s))^p}dN(s)
\eqno{(6)}
$$
for some positive parameters $c,k,p.$ 

This power law form for $\lambda(t)$ in (6) was applied in the geological model called Omori's law, and used to predict the rate of aftershocks caused by an earthquake. 

Many generalizations  of Hawkes processes have been proposed. 
They include, in particular, multi-dimensional Hawkes processes, non-linear Hawkes processes, mixed diffusion-Hawkes models, Hawkes models with shot noise exogenous events, Hawkes processes with generation dependent kernels.

\subsection{General Compound Hawkes Process (GCHP)}

{\bf Definition 7 (General Compound Hawkes Process (GCHP)).} Let $N(t)$ be any one-dimensional Hawkes process defined above. Let also $X_n$ be ergodic continuous-time finite (or possibly infinite but countable) state Markov chain, independent of $N(t),$ with space state $X,$ and $a(x)$ be any bounded and continuous function on $X.$ The general compound Hawkes process is defined as
$$
S_t=S_0+\sum_{k=1}^{N(t)}a(X_k).
\eqno{(7)}
$$  

\vspace{0.5cm}

{\bf Some Examples of GCHP}

\vspace{0.5cm}

{\bf 1. Compound Poisson Process}:
$
S_t=S_0+\sum_{k=1}^{N(t)}X_k,
$  
where $N(t)$ is a Poisson process and $a(X_k)=X_k$ are i.i.d.r.v.

\vspace{0.5cm}

{\bf 2. Compound Hawkes Process}: 
$
S_t=S_0+\sum_{k=1}^{N(t)}X_k,
$  
where $N(t)$ is a Hawkes process and $a(X_k)=X_k$ are i.i.d.r.v.

\vspace{0.5cm}

{\bf 3. Compound Markov Renewal Process}: 
$
S_t=S_0+\sum_{k=1}^{N(t)}a(X_k),
$  
where $N(t)$ is a renewal process and $X_k$ is a Markov chain.

\subsection{Risk Model based on General Compound Hawkes Process}

{\bf Definition 8 (RMGCHP: Finite State MC).} We define the risk model $R(t)$ based on GCHP as follows:
$$
R(t):=u+ct-\sum_{k=1}^{N(t)}a(X_k),
\eqno{(8)}
$$
where $u$ is the initial capital of an insurance company, $c$ is the rate of at which premium is paid, $X_k$ is continuous-time Markov chain in state space $X=\{1,2,...,n\},$ $N(t)$ is a Hawkes process, $a(x)$ is continuous and bounded function on X). $N(t)$ and $X_k$ are independent. 

{\bf Definition 8'. (RMGCHP: Infinite State MC).} We define the risk model $R(t)$ based on GCHP for infinite state but countable Markov chain as follows:
$$
R(t):=u+ct-\sum_{k=1}^{N(t)}a(X_k).
\eqno{(8')}
$$
Here: $X=\{1,2,...,n,...\}$-infinite but countable space of states for Markov chain $X_k.$

\vspace{0.5cm}

{\bf Some Examples of RMGCHP}

\vspace{0.5cm}

{\bf 1. Classical Risk Process (Cramer-Lundberg Risk Model)}: If $a(X_k)=X_k$ are i.i.d.r.v. and $N(t)$ is a homogeneous Poisson process, then $R(t)$ is a classical risk process also known as the Cramer-Lundberg risk model (see Asmussen and Albrecher (2010)). In the latter case we have compound Poisson process (CPP) for the outgoing claims. 

\vspace{0.5cm}

{\bf Remark 1.} Using this analogy, we call our risk process as a risk model based on general compound Hawkes process (GCHP).

\vspace{0.5cm}

{\bf 2. Risk Model based on Compound Hawkes Process}: If $a(X_k)=X_k$ are i.i.d.r.v. and $N(t)$ is a Hawkes process, then $R(t)$ is a risk process with non-stationary Hawkes claims arrival introduced in Stabile et al. (2010).

\section{LLN and FCLT for RMGCHP}

In this section we present LLN and FCLT for RMGCHP.

\subsection{LLN for RMGCHP}

{\bf Theorem 1 (LLN for RMGCHP)}. Let $R(t)$ be the risk model (RMGCHP) defined above in (8), and $X_k$ be an ergodic Markov chain with stationary probabilities $\pi^*_n.$ Then
$$
\lim_{t\to+\infty}\frac{R(t)}{t}=c-a^*\frac{\lambda}{1-\hat\mu},
\eqno{(9)}
$$

where $a^*=\sum_{k\in X}a(k)\pi_k^*,$ and $0<\hat\mu:=\int_0^{+\infty}\mu(s)ds<1.$

{\bf Proof.} (Follows from Swishchuk (2017) ('General Compound Hawkes Processes in Limit Order Books', working paper. Available on arXiv: \\
https://arxiv.org/submit/1929048)).

From (8) we have 
$$
R(t)/t=u/t+c-\sum_{i=1}^{N(t)}a(X_k)/t.
\eqno{(10)}
$$
The first term goes to zero when $t\to+\infty.$ 
From the other side, w.r.t. the strong LLN for Markov chains (see, e.g., Norris (1997))
$$
\frac{1}{n}\sum_{k=1}^{n}a(X_k)\to_{n\to+\infty}  a^*,
\eqno{(11)}
$$
where $a^*$ is defined in (9).

Finally, taking into account (10) and (11), we obtain:
$$
\sum_{i=1}^{N(t)}a(X_k)/t=\frac{N(t)}{t}\frac{1}{N(t)}\sum_{i=1}^{N(t)}a(X_k)\to_{t\to+\infty}a^*\frac{\lambda}{1-\hat\mu},
$$
and the result in (9) follows. 

We note, that we have used above the result that $N(t)/t\to_{t\to+\infty}\lambda/(1-\hat\mu).$ (See, e.g.,  Bacry, Mastromatteo and Muzy (2015) or Daley, D.J. and Vere-Jones, D. (1988)). Q.E.D.

\vspace{0.5cm}

{\bf Remark 2.} When $a(X_k)=X_k$ are i.i.d.r.v., then $a^*=EX_k.$

\vspace{0.5cm}

{\bf Remark 3.} When $\mu(t)=\alpha e^{-\beta t}$ is exponential, then $\hat\mu=\alpha/\beta.$

\subsection{FCLT for RMGCHP}

{\bf Theorem 2 (FCLT for RMGCHP).} Let $R(t)$ be the risk model (RMGCHP) defined above in (8), and $X_k$ be an ergodic Markov chain with stationary probabilities $\pi^*_n.$ Then 

$$
\lim_{t\to+\infty}\frac{R(t)-(ct-a^*N(t))}{\sqrt{t}}=^D\sigma\Phi(0,1),
\eqno{(12)}
$$
(or in Skorokhod topology (see Skorokhod (1965))

$$
\lim_{n\to+\infty}\frac{R(nt)-(cnt-a^*N(nt))}{\sqrt{n}}=\sigma W(t) )
\eqno{(12')}
$$
where $\Phi(\cdot,\cdot)$ is the standard normal random variable ($W(t)$ is a standard Wiener process), 
$$
\begin{array}{rcl}
\sigma&:=&\sigma^*\sqrt{\lambda/(1-\hat\mu)},\\
(\sigma^*)^2&:=&\sum_{i \in X} \pi^*_iv(i), \\
0<\hat\mu&:=&\int_0^{+\infty}\mu(s)ds<1,\\
\end{array}
\eqno{(13)}
$$

and 

$$
\begin{array}{rcl}
v(i)&=& b(i)^2\\
&+&\sum_{j\in X}(g(j)-g(i))^2P(i,j)-2b(i)\sum_{j\in\mathcal{S}}(g(j)-g(i))P(i,j),\\
b&=&(b(1),b(2),...,b(n))',\\
b(i):&=&a^*-a(i), \\
g:&=&(P+\Pi^*-I)^{-1}b,\\
a^*&:=&\sum_{i\in X}\pi^*_ia(i),\\
\end{array}
\eqno{(14)}
$$
$P$ is a transition probability matrix for $X_k,$, i.e., $P(i,j)=P(X_{k+1}=j|X_k=i),$ $\Pi^*$ denotes the matrix of stationary distributions of $P$ and $g(j)$ is the jth entry of $g.$

{\bf Proof.} (Follows from Swishchuk (2017) ('General Compound Hawkes Processes in Limit Order Books', working paper. Available on arXiv:\\
 https://arxiv.org/submit/1929048)). 
From (8) it follows that
$$
R(t)/\sqrt{t}=(u+ct-\sum_{i=1}^{N(nt)}a(X_k))/\sqrt{t},
$$
and
$$
R(t)/\sqrt{t}=(u+ct+\sum_{i=1}^{N(t)}(a^*-a(X_k))-N(t)a^*)/\sqrt{t},
\eqno{(15)}
$$
where $a^*$ is defined in (14)).

Therefore,
$$
\frac{R(t)-(ct-N(t) a^*)}{\sqrt{t}}=\frac{u+\sum_{i=1}^{N(t)}(a^*-a(X_k))}{\sqrt{t}}.
\eqno{(16)}
$$
As long as $\frac{u}{\sqrt{t}}\to_{t\to+\infty}0,$ we have to find the limit for 
$$\frac{\sum_{i=1}^{N(t)}(a^*-a(X_k))}{\sqrt{t}}$$ 
when $t\to+\infty.$

Consider the following sums
$$
R^*_n:=\sum_{k=1}^{n}(a(X_k)-\hat a^*)
\eqno{(17)}
$$
and
$$
U^*_n(t):=n^{-1/2}[(1-(nt-\lfloor nt\rfloor))R^*_{\lfloor nt\rfloor)}+(nt-\lfloor nt\rfloor))R^*_{\lfloor nt\rfloor)+1}],
\eqno{(18)}
$$
where $\lfloor\cdot\rfloor$ is the floor function.

Following the martingale method from Vadori and Swishchuk (2015), we have the following weak convergence in the Skorokhod topology (see Skorokhod (1965)):
$$
\hat U^*_n(t)\to_{n\to+\infty}\sigma^* W(t),
\eqno{(19)}
$$
where $\sigma^*$ is defined in (13).

We note again, that w.r.t LLN for Hawkes process $N(t)$ (see, e.g., Daley, D.J. and Vere-Jones, D. (1988)) we have:
$$
\frac{N(t)}{t}\to_{t\to+\infty}\frac{\lambda}{1-\hat\mu},
$$
or
$$
\frac{N(nt)}{n}\to_{n\to+\infty}\frac{t\lambda}{1-\hat\mu},
\eqno{(20)}
$$ 
where $\hat\mu$ is defined in (13).

Using change of time in (19), $t\to N(t)/t,$ we can find from (19) and (20):
$$
U^*_n(N(nt)/n)\to_{n\to+\infty}\sigma W\Big(t\lambda/(1-\hat\mu)\Big),
$$
or
$$
U^*_n(N(nt)/n)\to_{n\to+\infty}\sigma\sqrt{\lambda/(1-\hat\mu)}W(t),
\eqno{(21)}
$$
where $W(t)$ is the standard Wiener process, and $\sigma^*$ and $\hat\mu$ are defined in (13).
The result (12) now follows from (15)-(21). Q.E.D.

\vspace{0.5cm}

{\bf Remark 4.} When $a(X_k)=X_k\in \{+\delta,-\delta\}$ are independent and $P(1,2)=P(2,1)=\pi^*=1/2,$ then $a^*=0$ and $\sigma^*=+\delta.$

\vspace{0.5cm}

{\bf Remark 5.} When $a(X_k)=X_k\in \{+\delta,-\delta\}$ are independent and $P(1,2)=P(2,1)=p,$ then $\pi^*=1/2,$ $a^*=0$ and $(\sigma^*)^2=\delta^2p/(1-p).$

\vspace{0.5cm}

{\bf Remark 6.} When $a(X_k)=X_k\in \{+\delta,-\delta\}$ is two-state Markov chain and $P(1,1)=p', P(2,2)=p,$ then $a^*=\delta(2\pi^*-1)$ and 
$$
(\sigma^*)^2=4\delta^2(\frac{1-p'+\pi^*(p'-p)}{(p+p'-2)^2}-\pi^*(1-\pi^*).
$$

{\bf Remark 7.}  When $a(X_k)=X_k$ are i.i.d.r.v., then $\sigma^*=Var(X_k)$ in (13) and $\sigma=Var(X_k)\sqrt{\lambda/(1-\hat\mu)}.$
\section{Applications of LLN and FCLT for RMGCHP}

In this section we consider some applications of LLN and FCLT for RMGCHP that include net profit condition, premium principle and ruin and ultimate ruin probabilities.

\subsection{Application of LLN: Net Profit Condition}

From Theorem 1 (LLN for RMGCHP) follows that net profit condition has the following form:

\vspace{0.5cm}

{\bf Corollary 1 (NPC for RMGCHP).}
$$
c>a^*\frac{\lambda}{1-\hat\mu},
\eqno{(22)}
$$
where $a^*=\sum_{k\in X}a(k)\pi_k^*.$

\vspace{0.5cm}

{\bf Corollary 2 (NPC for RMCHP).} When $a(X_k)=X_k$ are i.i.d.r.v., then $a^*=EX_k,$ and the net profit condition in this case has the form
$$
c>\frac{\lambda}{1-\hat\mu}\times E[X_k].
$$

\vspace{0.5cm}

{\bf Corollary 3 (NPC for RMCPP).} Of course, in the case of Poisson process $N(t)$ ($\hat\mu=0$) we have well-known net profit condition:
$$
c>\lambda\times E[X_k].
$$

\subsection{Application of LLN: Premium Principle}

A premium principle is a formula for how to price a premium against an insurance risk. There many premium principles, and the following are three classical examples of premium principles ($S_t=\sum_{k=1}^{N(t)}a(X_k)$):

$\bullet$ The expected value principle: $c=(1+\theta)\times E[S_t]/t,$ \\
where the parameter $\theta>0$ is the safety loading;

$\bullet$ The variance principle: $c=E[S_t]/t+\theta\times Var[S_t/t];$

$\bullet$ The standard deviation principle: $c=E[S_t]/t+\theta\times \sqrt{Var[S_t/t]}.$

We present here the expected value principle as one of the premium principles (that follows from Theorem 1 (LLN for RMGCHP)):

\vspace{0.5cm}

{\bf Corollary 4 (Premium Principle for RMGCHP)}

$$
c=(1+\theta)\frac{a^*\lambda}{1-\hat\mu},
\eqno{(23)}
$$
where the parameter $\theta>0$ is the safety loading.

\subsection{Application of FCLT for RMGCHP: Ruin and Ultimate Ruin Probabilities}

\subsubsection{Application of FCLT for RMGCHP: Approximation of RMGCHP by a Diffusion Process}

From Theorem 2 (FCLT for RMGCHP) it follows that risk process $R(t)$ can be approximated by the following diffusion process $D(t):$
$$
R(t)\approx u+ct-N(t)a^*+\sigma W(t):=u+D(t),
$$
where $a^*$ and $\sigma$ are defined above, $N(t)$ is a Hawkes process and $W(t)$ is a standard Wiener process.

It means that our diffusion process $D(t)$ has drift $(c-a^*\lambda/(1-\hat\mu))$ and diffusion coefficient $\sigma,$ i.e., $D(t)$ is $N(c-a^*\lambda/(1-\hat\mu)t,\sigma^2t)$-distributed. 

We use the diffusion approximation of the RMGCHP to calculate the ruin probability in a finite time interval $(0,\tau).$ 

\subsubsection{Application of FCLT for RMGCHP: Ruin Probabilities}

The ruin probability up to time $\tau$ is given by ($T_u$ is a ruin time)

$$
\begin{array}{crl}
\vspace{1cm}
\psi(u,\tau)&=&1-\phi(u,\tau)=P(T_u<\tau)\\
\vspace{1cm}&=&P(\min_{0<t<\tau}R(t)<0)\\
&=&P(\min_{0<t<\tau}D(t)<-u).\\
\end{array}
$$

Applying now the result for ruin probabilities for diffusion process (see, e.g., Asmussen (2000) or Asmussen and Albrecher (2010)) we obtain the following

{\bf Theorem 3 (Ruin Probability for Our Diffusion Process)}:

$$
\begin{array}{crl}
\vspace{1cm}
\psi(u,\tau)&=&\Phi(-\frac{u+(c-a^*\lambda/(1-\hat\mu))\tau}{\sigma\sqrt{\tau}})\\
&+&e^{-\frac{2(c-a^*\lambda/(1-\hat\mu))}{\sigma^2}u}\Phi(-\frac{u-(c-a^*\lambda/(1-\hat\mu))\tau}{\sigma\sqrt{\tau}}),\\
\end{array}
\eqno{(24)}
$$

where $\Phi$ is the standard normal distribution function.

\subsubsection{Application of FCLT for RMGCHP: Ultimate Ruin Probabilities}

Letting $\tau\to+\infty$ in Theorem 3 above, we obtain:

\vspace{0.5cm}

{\bf Corollary 5 (The Ultimate Ruin Probability for RMGCHP)}:

$$
\psi(u)=1-\phi(u)=P(T_u<+\infty)=e^{-\frac{2(c-a^*\lambda/(1-\hat\mu))}{\sigma^2}u},
\eqno{(25)}
$$

where $\sigma$ and $\hat\mu$ are defined in Theorem 2 (FCLT for RMGCHP).

\subsection{Application of FCLT for RMGCHP: The Distribution of the Time to Ruin}

From Theorem 3 and Corollary 5 follows:

\vspace{0.5cm}

{\bf Corollary 6 (The Distribution of the Time to Ruin)}. The distribution of the time to ruin, given that ruin occurs is:

$$
\begin{array}{rcl}
\vspace{0.5cm}
\frac{\psi(u,\tau)}{\psi(u)}&=&P(T_u<\tau|T_u<+\infty)\\
\vspace{0.5cm}
&=&e^{\frac{2(c-a^*\lambda/(1-\hat\mu))}{\sigma^2}u}\Phi(-\frac{u+(c-a^*\lambda/(1-\hat\mu))\tau}{\sigma\sqrt{\tau}})\\
&+&\Phi(-\frac{u-(c-a^*\lambda/(1-\hat\mu))\tau}{\sigma\sqrt{\tau}})\\.
\end{array}
$$

Differentiation in previous distribution by $u$ gives the probability density function $f_{T_u}(\tau)$ of the time to ruin:

\vspace{0.5cm}

{\bf Corollary 7 (The Probability Density Function of the Time to Ruin)}:

$$
f_{T_u}(\tau)=\frac{u}{\sigma\sqrt{2\pi}}\tau^{-3/2}e^{-\frac{(u-(c-a^*\lambda/(1-\hat\mu))\tau)^2}{2\sigma^2\tau}},\quad \tau>0.
\eqno{(26)}
$$

\vspace{0.5cm}

{\bf Remark 8 (Inverse Gaussian Distribution)}: Substituting $u^2/\sigma^2=a$ and $u/(c-a^*\lambda/(1-\hat\mu))=b$ in the density function we obtain:

$$
f_{T_u}(\tau)=(\frac{a}{2\pi\tau^3})^{1/2}e^{-\frac{a}{2\tau}(\frac{\tau-b}{\sigma})^2}, \quad \tau>0,
$$

which is the standard Inverse Gaussian distribution with expected value $u/(c-a^*\lambda/(1-\hat\mu))$ and variance $u\sigma^2/(c-a^*\lambda/(1-\hat\mu)).$

\vspace{0.5cm}

{\bf Remark 9 (Ruin Occurs with $P=1$)}: If $c=a^*\lambda/(1-\hat\mu),$ then ruin occurs with $P=1$ and the density function is obtained from Corollary 6 with $c=a^*\lambda/(1-\hat\mu),$ i.e.,

$$
f_{T_u}(\tau)=\frac{u}{\sigma\sqrt{2\pi}}\tau^{-3/2}e^{-\frac{u^2}{2\sigma^2\tau}},\quad \tau>0.
$$

The distribution function is :
$$
F_{T_u}(\tau)=2\Phi(-\frac{u}{\sigma\sqrt{\tau}}), \quad \tau>0.
$$

\section{Applications of LLN and FCLT for \\ RMCHP}

In this section we list the applications of LLN and FCLT for risk model based on compound Hawkes process (RMCHP). The LLN and FCLT for RMCHP follow from Theorem 1 and Theorem 2 above, respectively. In this case $a(X_k)=X_k$ are i.i.d.r.v. and $a^*=EX_k,$ and our risk model $R(t)$ based on compound Hawkes process  $N(t)$ (RMCHP) has the following form:
$$
R(t)=u+ct-\sum_{k=1}^{N(t)}X_k,
$$ 
where $N(t)$ is a Hawkes process.

\subsection{Net Profit Condition for RMCHP}

From (22) it follows that net profit condition for RMCHP  has the following form ($a^*=EX_k$):
$$
c>\frac{\lambda EX_1}{1-\hat\mu}.
$$

\subsection{Premium Principle for RMCHP}

From (23) it follows that premium principle for RMCHP has the following  form:

$$
c=(1+\theta)\frac{\lambda EX_1}{1-\hat\mu},
$$
where $\theta>0$ is the safety loading parameter.

\subsection{Ruin Probability for RMCHP}

From (24) it follows that the ruin probability for RMCHP has the following form:
$$
\begin{array}{crl}
\vspace{1cm}
\psi(u,\tau)&=&\Phi(-\frac{u+(c-EX_1\lambda/(1-\hat\mu))\tau}{\sigma\sqrt{\tau}})\\
&+&e^{-\frac{2(c-EX_1\lambda/(1-\hat\mu))}{\sigma^2}u}\Phi(-\frac{u-(c-EX_1\lambda/(1-\hat\mu))\tau}{\sigma\sqrt{\tau}}).
\end{array}
$$

{\bf Remark 10.} Here, $\sigma=Var(X_k)\sqrt{\lambda/(1-\hat\mu)}$ (see Remark 7.).

\subsection{Ultimate Ruin Probability for RMCHP}

From (25) it follows that the ultimate ruin probability for RMCHP has the following form:
$$
\psi(u)=1-\phi(u)=P(T_u<+\infty)=e^{-\frac{2(c-EX_1\lambda/(1-\hat\mu))}{\sigma^2}u}.
$$

\subsection{The Probability Density Function of the Time to Ruin}

From (26) it follows that the probability density function of the time to ruin for RMCHP has the following form:
$$
f_{T_u}(\tau)=\frac{u}{\sigma\sqrt{2\pi}}\tau^{-3/2}e^{-\frac{(u-(c-EX_1\lambda/(1-\hat\mu))\tau)^2}{2\sigma^2\tau}},\quad \tau>0.
$$

\section{Applications of LLN and FCLT for \\ RMCPP}

In this section we list, just for completeness, the applications of LLN and FCLT for risk model based on compound Poisson process (RMCPP). The LLN and FCLT for RMCPP follow from Section 5 above. In this case $a(X_k)=X_k$ are i.i.d.r.v. and $a^*=EX_k,$ and $\hat\mu=0$ and our risk model $R(t)$ based on compound Poisson process $N(t)$ (RMCHP) has the following form:
$$
R(t)=u+ct-\sum_{k=1}^{N(t)}X_k,
$$ 
where $N(t)$ is a Poisson process.

Of course, all the results below are classical and well-known (see, e.g., Asmussen (2000)), and we list them just to show that they are followed from our results above.

\subsection{Net Profit Condition for RMCPP}

From (22) it follows that net profit condition for RMCPP has the following form ($a^*=EX_k$):
$$
c>\lambda EX_1.
$$

\subsection{Premium Principle for RMCPP}

From (23) it follows that premium principle for RMCPP has the following  form:

$$
c=(1+\theta)\lambda EX_1,
$$
where $\theta>0$ is the safety loading parameter.

\subsection{Ruin Probability for RMCPP}

From (24) it follows that the ruin probability for RMCPP has the following form:
$$
\begin{array}{crl}
\vspace{1cm}
\psi(u,\tau)&=&\Phi(-\frac{u+(c-EX_1\lambda)\tau}{\sigma\sqrt{\tau}})\\
&+&e^{-\frac{2(c-EX_1\lambda)}{\sigma^2}u}\Phi(-\frac{u-(c-EX_1\lambda)\tau}{\sigma\sqrt{\tau}}).
\end{array}
$$

{\bf Remark 11.} Here, $\sigma=Var(X_k)\sqrt{\lambda}$ because $\hat\mu=0$ (see remark 7.).

\subsection{Ultimate Ruin Probability for RMCPP}

From (25) it follows that the ultimate ruin probability for RMCPP has the following form:
$$
\psi(u)=1-\phi(u)=P(T_u<+\infty)=e^{-\frac{2(c-EX_1\lambda)}{\sigma^2}u}.
$$

\subsection{The Probability Density Function of the Time to Ruin for RMCPP}

From (26) it follows that the probability density function of the time to ruin for RMCPP has the following form:
$$
f_{T_u}(\tau)=\frac{u}{\sigma\sqrt{2\pi}}\tau^{-3/2}e^{-\frac{(u-(c-EX_1\lambda)\tau)^2}{2\sigma^2\tau}},\quad \tau>0.
$$

\section{Conclusion and Future Work} In this paper, we introduced a new model for the risk process based on general compound Hawkes process (GCHP) for the arrival of claims. We call it risk model based on general compound Hawkes process (RMGCHP). The Law of Large Numbers (LLN) and the Functional Central Limit Theorem (FCLT) have been proved. We also studied the main properties of this new risk model,  net profit condition, premium principle and ruin time (including ultimate ruin time) applying the LLN and FCLT for the RMGCHP. We showed similar results for risk model based on compound Hawkes process (RMCHP) and applied them to the classical risk model based on compound Poisson process (RMCPP). The future work will be devoted to the implementations of the obtained results to some insurance problems and preparation of numerical results.

\section*{Acknowledgement} The author thanks the organizers of the 21st International Congress on Insurance: Mathematics and Economics-IME 2017, July 3-5, 2017, TUW, Vienna, for their kind invitation to present the results of the paper. 

\section*{References}

\hspace{0.5cm}

Asmussen, S. (2000): {\it Ruin Probabilities}. World Scientific, Singapore.

Asmussen, S. and Albrecher, H. (2010): {\it Ruin Probabilities}. 2nd edition, World Scientific, Singapore.

Bacry, E., Mastromatteo, I. and Muzy, J.-F. (2015): Hawkes processes in finance. {\it Market Microstructure and Liquidity}, June 2015, Vol. 01, No. 012. 

Daley, D.J. and Vere-Jones, D. (1988): {\it An Introduction to the theory of Point Processes}. Springer.

Dassios, A. and Zhao, HB. (2011): A dynamic contagion process. {\it Adv. in Appl. Probab}, Volume 43, Number 3, 814-846.

Dassios, A. and Jang, J. (2012): A Double Shot Noise Process and Its Application in Insurance. {\it J. Math. System Sci.}, 2, 82-93.

Embrechts, P., Liniger, T. and Lin, L. (2011): Multivariate Hawkes Processes: An Application to Financial Data. {\it Journal of Applied Probability}, 48A, 367-378.

Hawkes, A. G. (1971): Spectra of some self-exciting and mutually exciting point processes. {\it Biometrika}, 58, 83-90.

Jang, J. and Dassios, A. (2013): A bivariate shot noise self-exciting process for insurance. {\it Insurance: Mathematics and Economics}, 53 (3), 524-532.

Norris, J. R. (1997): Markov Chains. In Cambridge Series in Statistical and Probabilistic Mathematics. UK: Cambridge University Press.

Skorokhod, A. (1965): Studies in the Theory of Random Processes. {\it Addison-Wesley}, Reading, Mass.,
(Reprinted by Dover Publications, NY).

Stabile, G. and G. L. Torrisi. (2010): Risk processes with non-stationary Hawkes arrivals. {\it Methodol. Comput. Appl. Prob.}, 12, 415-429.

Swishchuk A.V. and Goncharova S.Y. (1998): Optimal control of semi-Markov risk processes. {\it Nonlinear Oscillations}, No2, 122-131.

Swishchuk, A. (2000): {\it Random Evolutions and Their Applications: New Trends}. Kluwer AP, Dordrecht.

Swishchuk, A. (2017): Risk model based on compound Hawkes process. Abstract, IME 2017, Vienna.

Swishchuk, A. (2017): General Compound Hawkes Processes in Limit Order Books. {\it Working Paper}, U of Calgary, 32 pages, June 2017. Available on arXiv: https://arxiv.org/submit/1929048

Swishchuk, A., Chavez-Casillas, J., Elliott, R. and Remillard, B. (2017): Compound Hawkes processes in limit order books. Available on SSRN: 
https://papers.ssrn.com/sol3/papers.cfm?abstract\_id=2987943.

Vadori, N. and Swishchuk, A. (2015): Strong law of large numbers and central limit theorems for functionals of inhomogeneous Semi-Markov processes. {\it Stochastic Analysis and Applications}, 13 (2), 213?243.

\end{document}